\documentclass[11pt,a4paper]{article}
\pdfoutput=1
\usepackage{jheppub} 

\usepackage{amsmath} 
\usepackage{array,relsize,float}
\usepackage{pstricks}
\usepackage{slashed}
\usepackage{tikz-cd} 
\usepackage{multicol} % Allows for multiple columns
\usepackage{pdfpages}
\usepackage{array}
\usepackage{tikz-cd}
\usepackage{amsfonts,layout,appendix,subfigure}
\allowdisplaybreaks
\input{epsf} 
\def\rar{\rightarrow}

\def\eqv{\equiv}
\def\nn{\nonumber}

\def\mb{\mathbf}

\def\eps{\epsilon}

\def\wt{\widetilde}
\def\fr{\frac}
\def\kotm{\tilde{k}_1^{\mu}}
\def\kotn{\tilde{k}_1^{\nu}}
\def\kttm{\tilde{k}_2^{\mu}}
\def\kttn{\tilde{k}_2^{\nu}}
\def\kthtm{\tilde{k}_3^{\mu}}
\def\kthtn{\tilde{k}_3^{\nu}}
\renewcommand{\thefigure}{\arabic{figure}}

\def\beq{\begin{equation}} \def\eeq{\end{equation}}
\def\beqn{\begin{eqnarray}} \def\eeqn{\end{eqnarray}}
\def\bom#1{{\mbox{\boldmath $#1$}}} \def\to{\rightarrow}
\def\nn{\nonumber}

\def\Eq#1{eq.~(\ref{#1})}

\def\beq{\begin{equation}}
\def\eeq{\end{equation}}
\def\bea{\begin{eqnarray}}
\def\eea{\end{eqnarray}}
\def\beqn{\begin{eqnarray}} \def\eeqn{\end{eqnarray}}
\def\beeq{\begin{eqnarray}}
\def\eeeq{\end{eqnarray}}

\def\ep{\epsilon}
\def\eps{\varepsilon}
\def\nn{\nonumber}
\def\Eq#1{eq.~(\ref{#1})}

\newcommand{\la}{\langle}
\newcommand{\ra}{\rangle}
\def\bom#1{{\mbox{\boldmath $#1$}}}
\def\sp{{\bom {Sp}}}
\def\ket#1{|{#1}\ra}
\def\bra#1{\la{#1}|}

\def\wp{\widetilde P}

\newcommand\as{\alpha_{\mathrm{S}}}

\def\as{\alpha_{\rm S}}

\pdfoutput=1

\def\bt{\boldsymbol{t}}

\newcommand{\valencia}{Instituto de F\'{\i}sica Corpuscular,\\ Universitat de Val\`{e}ncia -- Consejo Superior de Investigaciones Cient\'{\i}ficas,\\ Parc Cient\'{\i}fic, E-46980 Paterna, Valencia, Spain}
\newcommand{\europea}{Escuela de Ciencias, Ingenier\'ia y Diseño, Universidad Europea de Valencia,\\ Paseo de la Alameda 7, 46010 Valencia, Spain}
\newcommand{\salamanca}{Departamento de F\'isica Fundamental e IUFFyM, Universidad de Salamanca,\\ Plaza de la Merced S/N, 37008 Salamanca, Spain}
\begin{document}
\title{Triple-collinear splittings with massive particles}

\author[a]{Prasanna K. Dhani,}
\author[a]{Germ\'an Rodrigo,}
\author[b,c]{and German~F. R.~Sborlini} 
\affiliation[a]{\valencia}
\affiliation[b]{\salamanca}
\affiliation[c]{\europea}

\emailAdd{dhani@ific.uv.es}
\emailAdd{german.rodrigo@csic.es}
\emailAdd{german.sborlini@usal.es}

\preprint{IFIC/23-46}

\abstract{We analyze in detail the most singular behaviour of processes involving triple-collinear splittings with massive particles in the quasi-collinear limit,
and present compact expressions for the splitting amplitudes and the corresponding splitting kernels at the squared-amplitude level. Our expressions fully agree with well-known triple-collinear splittings in the massless limit, which are used as a guide to achieve the final expressions. These results are important to quantify dominant mass effects in many observables, and constitute an essential ingredient of current high-precision computational frameworks for collider phenomenology.}

\keywords{Factorization, Renormalization Group, Higher-Order Perturbative Calculations, Quark Masses, Resummation}
\arxivnumber{2310.05803}
\setcounter{page}{1}
\maketitle

%***********************************************************************************************************************************************%
\section{Introduction}
\label{sec:introduction}
Owing to the non-Abelian nature of Quantum Chromodynamics (QCD) due to multi-gluon interactions, a very relevant topic of study is the singular structure of scattering amplitudes in various infrared (IR), soft and collinear, limits. The soft limit is achieved when one or more gluons are emitted in  a hard-scattering process with vanishingly small energies and the collinear limit is achieved when two or more partons, quarks and gluons, are emitted parallel to each other i.e. with vanishingly small angles of separation. In these IR limits, the scattering amplitude for the process becomes singular and it exhibits a factorized structure~\cite{Collins:1989gx, Catani:2011st,Forshaw:2012bi}. The singular behaviour is given by the QCD factorization formulae \cite{Collins:1989gx} and it is captured by factors which have minimal dependence on the hard-scattering process under consideration.

In quest for ever increasing precise theoretical predictions for observables at high-energy colliders such as the CERN's Large Hadron Collider (LHC), the framework of perturbation theory plays a pivotal role. To this end, a higher order theory prediction requires to combine contributions from both virtual loop and real radiation Feynman diagrams for a  meaningful comparison with the experimental data. Virtual loop diagrams in the IR region lead to singular configurations in four space-time dimensions. Similarly, real radiation of soft and/or collinear partons produce kinematical singularities after the phase space integration over the emitted partons. In the context of dimensional regularization (DR) \cite{Bollini:1972ui,tHooft:1972tcz,Cicuta:1972jf} where one regulates these singularities by performing the computations in arbitrary space-time dimensions, the understanding of these IR singular configurations and its isolation for the construction of efficient methods to evaluate cross sections rely on the general factorization properties of hard-scattering QCD amplitudes in both soft and collinear limits. Based on this understanding, general algorithms for the computation of IR finite jet cross sections at next-to-leading order (NLO) were proposed through celebrated works in refs.~\cite{Catani:1996vz,Frixione:1995ms}. Going forward to the next perturbative orders i.e. next-to-NLO (NNLO) and next-to-NNLO (N3LO), in the last twenty years or so, a gigantic effort has been put forth and a variety of different techniques\footnote{See the review in refs.~\cite{TorresBobadilla:2020ekr,Caola:2022ayt} and references therein.} now exist with the ability to make finite predictions. All of these methods rely on the knowledge of underlying QCD factorization of hard-scattering amplitudes in IR limits\footnote{For a review on IR structure of gauge theories, see ref.~\cite{Agarwal:2021ais} and references therein.}.

IR divergences emerging from phase space integration and virtual loop integration cancel between themselves for a sufficiently inclusive  observable (there still remains collinear divergences in the presence of identified partons and they are responsible for the evolution of non-perturbative parton distribution functions (PDFs) in the case of initial states and fragmentation functions (FFs) in the case of final states). However, close to the boundary of phase space, real and virtual loop contributions are strongly imbalanced due to differences in available phase space. As a result, the mechanism of IR pole cancellation leaves residual effects in the form of large logarithmic contributions for a wide class of hard-scattering observables. To obtain a reliable perturbative prediction, these large logarithmic contributions have to be unveiled to high powers in QCD strong coupling $\as (\equiv g_{\rm S}^2/(4\pi))$, and possibly resummed to all orders in perturbation theory. Soft and collinear singular factors from the QCD factorization formulae play a crucial role in achieving this goal through the computation of observable dependent and process independent integrated quantities. For instance, a novel methodology was recently developed in refs.~\cite{Catani:2022sgr, Dhani:2022dii} to compute logarithmically-enhanced contributions of collinear origin for the resummation in transverse momentum and N-jettiness distributions. Needless to say, soft and collinear factorization formulae form the basis of parton shower methods which are implemented in Monte Carlo event generators to describe the exclusive structure of hadronic final states\footnote{Several examples of parton shower generators can be found in the review ref. \cite{Hoche:2014rga}, and references therein.}.

Among IR factorization formulae, collinear factorization~\cite{Bern:1993qk,Bern:1994zx,Bern:1995ix,Kosower:1999xi,Bern:1999ry,Catani:1999ss,Catani:2003vu,Catani:2011st} is understood to be universal for theory predictions at colliders. Precisely, the divergent factors called splitting matrices at the amplitude level and splitting kernels at the squared amplitude level, only depend on the momenta and quantum numbers of those external partons involved in the collinear splitting process. In other words, splitting matrices and splitting kernels do not depend on the specific hard-scattering process under consideration and its production mechanism\footnote{Still, non-Abelian interactions at higher orders might introduce correlations among collinear and non-collinear states, breaking strict collinear factorization, as discussed in refs.~\cite{Catani:2011st,Forshaw:2012bi}.}.

In massless QCD, i.e. QCD with $N_f=5$ massless flavors, tree-level double-parton collinear limits for various splitting processes at $\mathcal{O}(\as)$ are well known~\cite{Altarelli:1977zs} and they are proportional to the real radiation contribution to the Altarelli-Parisi kernels for the LO evolution of the PDFs. At $\mathcal{O}(\as^2)$, both  triple-parton collinear limit at tree level~\cite{Campbell:1997hg,Catani:1998nv,Catani:1999ss} and double-parton collinear limit at one-loop level~\cite{Bern:1993qk, Bern:1994zx,Bern:1995ix,Bern:1998sc,Bern:1999ry,Kosower:1999rx,Catani:2011st,Sborlini:2013jba} are fully known. The contributions to various splitting processes at $\mathcal{O}(\as^3)$ (and beyond) are computed in refs.~\cite{DelDuca:1999iql,Birthwright:2005ak,Birthwright:2005vi,Delto:2019asp,DelDuca:2020vst,Catani:2003vu,Catani:2011st,Sborlini:2014mpa,Sborlini:2014kla,Badger:2015cxa,Czakon:2022fqi,Bern:2004cz,Badger:2004uk,Dixon:2019lnw,Vogt:2004mw,Vogt:2005dw,Moch:2004pa,Hoche:2017iem,Snigirev:2016uaq}.

While there has been a lot of progress in understanding the collinear factorization of hard-scattering matrix elements with massless partons, the case with massive partons needs more dedicated efforts in view of modern day precision colliders. The production of heavy flavors such as bottom and charm quarks at high-energy colliders has been a topic of interest due to their relevance in Standard Model (SM) and beyond the SM physics. For better theory predictions, it is crucial to consider higher-order contributions beyond the LO in perturbation theory. To this end, it is also important to have control over explicit IR poles and possibly, large logarithmic contributions as a result of scale hierarchies. QCD factorization with massive particles plays a big role in achieving these goals.

It is well known that radiation from massive partons in the collinear limit is suppressed, which constitutes the so-called dead-cone effect~\cite{Dokshitzer:1991fd}, and the appearance of explicit collinear poles in perturbative computations is screened due to the presence of mass. However, they can lead to IR finite contributions proportional to ln$^n\left(Q^2/m^2\right)$, where $Q$ is the typical scale of the hard-scattering process and $m$ is the mass of the heavy parton. In the kinematical region where $Q^2\gg m^2$, these large logarithmic contributions can spoil both the numerical convergence and reliability of theoretical prediction of a generic observable. The good news is that these contributions can be computed in a process independent manner and resummed to all orders in perturbation theory as they are related to singular behaviour of matrix elements in the $m\rightarrow 0$ limit. The singular behaviour is controlled by the QCD factorization formulae in the {\it quasi-collinear} limit~\cite{Catani:2000ef,Keller:1998tf}, in the same way that the IR divergences are controlled by the soft and collinear factorization formulae.

At $\mathcal{O}(\as)$, double-parton quasi-collinear limit for various tree-level splitting processes are computed long back in refs.~\cite{Catani:2000ef,Keller:1998tf}. Consequently, general algorithms~\cite{Dittmaier:1999mb,Roth:1999kk,Phaf:2001gc,Catani:2002hc} were proposed analogously to the case of massless partons~\cite{Catani:1996vz,Frixione:1995ms} for the computation of jet observables. In this paper, we take the first step in extending the QCD factorization in quasi-collinear limit to $\mathcal{O}(\as^2)$ and we study simultaneous collinear limit of three partons in sight of the importance for accurately predicting dominant mass effects in phenomenological analysis. We compute the complete set of splitting matrices at the amplitude level as well as splitting kernels at the squared amplitude level in arbitrary space-time dimensions, $d=4-2\epsilon$, and keep its dependence exact. At the squared amplitude level, we also fully take into account azimuthal correlations, specific to the gluon channel, which are necessary to construct some general algorithms to perform exact NNLO computations of jet cross sections involving massive particles.

The outline of the paper is the following. We start in section \ref{sec:formalism} by reviewing the basic concepts of collinear factorization and provide a formal definition of the quasi-collinear limit. In particular, we provide relevant details about the calculation of splitting amplitudes and polarized/unpolarized splitting kernels with massive particles. Then, in section \ref{sec:doublesplittings}, we review the available results for the double-collinear splitting processes with massive partons in QCD. After that, we provide explicit compact expressions for the QCD triple-collinear splitting amplitudes and splitting kernels with massive partons in sections \ref{sec:triplesplittings} and~\ref{sec:SquaredTriple}, respectively. We consider all the possible tree-level triple splitting processes, thus having access to all the required splitting amplitudes and splitting kernels for a complete double-real collinear contribution to a NNLO cross-section calculation with massive partons. In addition, we explain how to recover the corresponding QED results. Finally, in section \ref{sec:conclusions}, we present the conclusions and discuss potential applications of our results to high-precision QCD phenomenology.

%***********************************************************************************************************************************************%
\section{Multiple collinear and quasi-collinear limits of scattering amplitudes}
\label{sec:formalism}
Let us briefly describe the multiple quasi-collinear limit of scattering amplitudes at tree-level, and introduce various concepts and definitions. We consider a generic hard-scattering process in QCD with massive quarks and anti-quarks, that involves $n$ final-state partons along with any number of non-coloured particles such as Higgs bosons, electroweak bosons, photons etc. We denote as ${\cal M}^{(0)}\left(p_1, \ldots, p_m, \ldots,  p_n\right)$ the corresponding tree-level matrix element, with $p_i^{\mu}$ being the four-momentum associated with $i^{\rm th}$-particle, while the dependence on additional non-QCD particles through their momenta and quantum numbers is always understood. We denote the flavor of parton $i$ with $a_i$ and we will avoid indicating the flavor indices explicitly unless they are strictly required.

The external partons $a_i$ are on-shell i.e. $p_i^2=m_i^2$. Also, note that we always define the momenta $p_i$'s as outgoing, but the time component $p_i^0$ of the four-momentum vector $p_i^{\mu}$ is not constrained to be positive-definite. Different types of physical processes are represented by applying crossing symmetry to the same matrix element ${\cal M}^{(0)}$ (for brevity, we also avoid indicating the dependence on momenta unless it is necessary). For instance, if $p_i$ has positive energy, ${\cal M}^{(0)}$ describes a physical process that produces the parton $i$ in the final state; if $p_i$ has negative energy, the matrix element ${\cal M}^{(0)}$ describes a physical process produced by the collision of anti-parton $\bar{i}$ in the initial state.

We are interested in analysing the behaviour of the hard-scattering matrix element ${\cal M}^{(0)}$ in the kinematical configuration where $m$ out of $n$ external parton momenta become collinear to each other. Without any loss of generality, we assume them to be $p_1,\ldots,p_m$ and we define their sum to be $p_{1\ldots m} = p_1+\ldots +p_m$. The mass of the parent decaying particle with flavor $a$, which undergoes the splitting, is denoted by $m_{1\ldots m}$. Since, the parent decaying particle is generally off-shell ($p_{1\dots m}^2 \neq m_{1\ldots m}^2$), we define an on-shell vector
\begin{align}
    \wp^{\mu} = p_{1\dots m}^{\mu}-\frac{p_{1\ldots m}^2-m_{1\ldots m}^2}{2\, n \cdot p_{1\ldots m}}n^{\mu} \, ,
    \label{eq:Ptildegeneral}
\end{align}
such that it fulfils $\wp^2 = m_{1\ldots m}^2$. 
Then, we introduce the Sudakov parametrization for the collinear momenta~\cite{Catani:2002hc,Catani:2011st} as follows
\begin{align}
    p_i^{\mu} = x_i \wp^{\mu} + k_{\perp i}^{\mu} - \fr{k_{\perp i}^2+x_i^2m_{1\ldots m}^2-m_i^2}{x_i}\frac{n^{\mu}}{2\,n\cdot \wp}\ , \qquad i \in C\equiv\{1,\ldots,m\},
\end{align}
where the vector $\wp$ denotes the collinear direction and the introduction of an auxiliary light-like vector $n^{\mu}$ ($n^2=0$) is necessary to specify how the collinear direction is approached or, equivalently, to specify the transverse components $k_{\perp i}$ ($k_{\perp i}\cdot \wp =k_{\perp i}\cdot n= 0$, with $k_{\perp i}^2<0$).

Since, the collinear limit is invariant under longitudinal boosts along the direction of total momentum $p_{1\ldots m}$, we define following boost invariant kinematic variables
\begin{align}
\label{eq:zi}
z_i &= \fr{x_i}{\sum_{j=1}^{m} x_j} ,
\\
\label{eq:kti}
\tilde{k}_i^{\mu} &= k_{\perp i}^{\mu}-z_i\sum_{j=1}^{m} k_{\perp j}^{\mu}~.
\end{align}
Note that these boost invariant variables automatically satisfy the constraints $\sum_{i=1}^m z_i = 1$ and  $\sum_{i=1}^m \tilde{k}_{\perp i}=0$ and hence only $2(m-1)$ of them are independent. In terms of the variables introduced in eqs.~(\ref{eq:zi}) and (\ref{eq:kti}), the sub-energies and the dot products are expressed in the following way
\begin{align}
    s_{ij} &\eqv (p_i+p_j)^2 = -z_i z_j\left( \fr{\tilde{k}_j}{z_j}-\frac{\tilde{k}_i}{z_i}\right)^2 + (z_i+z_j)\left(\frac{m_i^2}{z_i}+\frac{m_j^2}{z_j} \right)~,
    \label{eq:sijmassive}
\\    
    \tilde{s}_{ij}&\eqv 2p_i\cdot p_j = z_i z_j\left[-\left( \fr{\tilde{k}_j}{z_j}-\frac{\tilde{k}_i}{z_i}\right)^2 + \fr{m_i^2}{z_i^2} + \fr{m_j^2}{z_j^2} \right]~,
    \label{eq:pijmassive}
\end{align}
and they are equivalent in the massless limit. 

Quasi-collinear limit~\cite{Catani:2000ef,Keller:1998tf} is reached when the transverse momenta $k_{\perp}$ is of $\mathcal{O}(m)$ or small. It turns out that, by virtue of eqs.~(\ref{eq:sijmassive}) and (\ref{eq:pijmassive}), the invariant mass of the splitting system depends on $\{k_{\perp i}^2,m_i^2\}$ and $m_{1\ldots m}^2$. So, this allows us to formally define the {\it quasi-collinear} kinematical region by uniformly rescaling these variables, i.e.
\beq
\label{eq:quasicollimit}
m_i \to \rho \, m_i~, \qquad k_{\perp i} \to \rho \, k_{\perp i}~, \qquad m_{1\ldots m} \to \rho \, m_{1\ldots m}~,
\eeq
and taking the limit $\rho \to 0$. Keeping the most singular terms in $\rho$ and neglecting sub-leading terms, we obtain the following amplitude level factorization formula
\beq
\ket{{\cal M}^{(0)}\left(p_1, \ldots,p_m,\ldots, p_n\right)} \simeq 
\sp^{(0)}_{a_1 \dots a_m}(p_1 , \ldots, p_m; \wp) \, 
\ket{\overline{{\cal M}}^{(0)}(\wp, p_{m+1}, \ldots, p_n)} \, , 
\label{eq:Factorization}
\eeq
where the factor $\sp^{(0)}_{a_1 \dots a_m}(p_1 , \ldots, p_m; \wp)$ is a matrix in color + spin space, called the splitting matrix that fully embodies the leading singular behaviour when the parent particle $a$ with momentum $p_{1\ldots m}$ undergoes collinear splitting into the particles $\{a_i\}_{i\in C}$, and it only depends on information carried by the quasi-collinear particles. In eq.~(\ref{eq:Factorization}), $\overline{{\cal M}}^{(0)}$ is called the reduced matrix element and it is obtained from the original amplitude by replacing $m$ collinear partons by the single parent parton of momentum $\wp$. Taking the massless limit of \Eq{eq:Factorization}, we obtain the well-known collinear factorization formula~\cite{Catani:2011st}. It is useful to define the colour-projected splitting matrices, and we have
\beq
Sp_{a_1\ldots a_m}^{(0)(c_1,\ldots, c_m;c)}\left(p_1,\ldots, p_m;\wp\right) \equiv
\langle c_1,\ldots,c_m|\sp_{a_1\ldots a_m}^{(0)}\left(p_1,\ldots, p_m;\wp\right) | c\rangle \, ,
\label{eq:ColorDECO}
\eeq
where $c$ is the colour of the parent parton and $c_1,\ldots,c_m$ denote the colour indices of the collinear partons.

An important observation related to \Eq{eq:Factorization} is that, the parent parton in the exact collinear or quasi-collinear limits is on-shell and it carries only physical degrees of freedom. In this way, the original scattering amplitude factorizes into two factors, namely the splitting matrix and the reduced matrix element, both containing {\it physical states} as external particles. Notice that this is valid both for massive and massless particles. The factorization formula in \Eq{eq:Factorization} is gauge independent, however, in order to unveil the splitting amplitude directly we need to work in a {\it physical gauge}. Explicitly, in physical gauges we have the following relations
\beqn
\frac{\imath \, d_{\mu \nu}(p,n)}{p^2 + \imath 0} &\to& \frac{\imath }{p^2+\imath 0} \, \sum_{\lambda \ {\rm phys.}} \, \varepsilon_\mu(\wp,\lambda) \varepsilon^*_\nu(\wp,\lambda)
\label{eq:polsumGLUON}
\eeqn
for gluons using the polarization tensor $d_{\mu\nu}(p,n)$, and
\beqn
\frac{\imath \, (\gamma^\mu p_\mu + m)}{p^2 - m^2 + \imath 0} &\to& \frac{\imath }{p^2-m^2+\imath 0} \, \sum_{\lambda \ {\rm phys.}} \, u(\wp,\lambda) \bar{u}(\wp,\lambda)
\label{eq:polsumQUARK}
\eeqn
for quarks, where $\lambda$ denotes the polarization (or spin).

Besides the splitting matrices, we are also interested in the polarized and unpolarized splitting kernels, which govern the collinear and quasi-collinear limits of the squared matrix elements. In order to define the polarized kernels, let us start from \Eq{eq:Factorization} and consider a gluon-initiated splitting process. Stripping the polarization vector from the splitting amplitude, we have
\beqn
\nonumber \ket{{\cal M}^{(0)}\left(p_1, \ldots,p_m,\ldots, p_n\right)} \simeq && \sum_{\lambda \, {\rm phys.}} \, \sp^{(0),\mu}_{a_1 \dots a_m}(p_1 , \ldots, p_m; \wp) \, \varepsilon_{\mu}(\wp,\lambda)\,
\\ && \times  \ket{{\cal M}^{(0)}(\wp^{-\lambda}, p_{m+1}, \ldots, p_n)} \, , 
\label{eq:FactorizationDER1}
\eeqn
and computing the square, we obtain
\beqn
\nonumber |{\cal M}^{(0)}\left(p_1, \ldots,p_m,\ldots, p_n\right)|^2 \simeq && \sum_{\lambda,\lambda' \, {\rm phys.}} \, \bra{{\cal M}^{(0)}(\wp^{-\lambda'}, p_{m+1}, \ldots, p_n)}\, \varepsilon_{\mu}(\wp,\lambda)\varepsilon^*_{\nu}(\wp,\lambda') \, 
\\ \nonumber && \times \left[\sp^{(0),\nu}_{a_1 \dots a_m}(p_1 , \ldots, p_m; \wp) \right]^{\dagger} \sp^{(0),\mu}_{a_1 \dots a_m}(p_1 , \ldots, p_m; \wp) \, \,
\\ && \times  \ket{{\cal M}^{(0)}(\wp^{-\lambda}, p_{m+1}, \ldots, p_n)} \, .
\label{eq:FactorizationDER2}
\eeqn
The second line of eq.~(\ref{eq:FactorizationDER2}) contains the product of two amputated splitting amplitudes, which leads us to define
\beq
\hat{P}^{(0), \, \mu \nu}_{a_1 \dots a_m} \equiv \left( \frac{s_{1 \ldots m}-m_{1\ldots m}^2}{8\pi\mu_0^{2\ep}\as} \right)^{m-1} \left[\sp^{(0),\nu}_{a_1 \dots a_m}(p_1 , \ldots, p_m; \wp) \right]^{\dagger} \sp^{(0),\mu}_{a_1 \dots a_m}(p_1 , \ldots, p_m; \wp) \, ,
\label{eq:PolarizedDEF}
\eeq
that corresponds to the polarized splitting kernel, with the proper normalization and $\mu_0$ denotes the DR scale. Note that analogous to the splitting matrix $\sp^{(0)}$ at the amplitude level, the splitting kernel $\hat{P}^{(0)}$ is a matrix in color + spin space and it encodes the leading singular behaviour of the squared amplitude in the quasi-collinear limit. In eq.~(\ref{eq:PolarizedDEF}), we have introduced $s_{1 \ldots m}\equiv p_{1\ldots m}^2$ and summed over final-state colours and spins, averaged over only colours of the parent decaying parton. By averaging over spins/polarizations of the parent parton, we obtain the $m$-parton unpolarized splitting kernel 
$\la \hat{P}^{(0)}_{a_1\ldots a_m} \ra$, which is a generalization of the customary 
Altarelli-Parisi LO splitting function. 
Explicitly, by fixing the normalization, the tree-level unpolarized splitting kernels are given by
\beq
\label{p0def}
\la \hat{P}_{a_1 \cdots a_m}^{(0)} \ra = 
\left( \frac{s_{1 \ldots m}-m_{1\ldots m}^2}{8\pi\mu_0^{2\ep}\as} \right)^{m-1} \,
{\overline {| \sp_{a_1 \dots a_m}^{(0)} |^2}}~.
\eeq
In the quark-initiated case, we have the relation
\beq
\la \hat{P}_{a_1 \cdots a_m}^{(0)} \ra   = \frac{1}{2} \, \delta^{s s'} \,\hat{P}^{(0), \, s s'}_{a_1 \dots a_m}  \, ,
\label{eq:quarkavg}
\eeq
because of helicity conservation. The spin indices $s$ and $s^\prime$ are of the parent collinear quark (anti-quark) in the amplitude and conjugate amplitude, respectively. In the gluon-initiated splitting process, the unpolarized kernel is obtained by contracting the polarized one with the polarization tensor of the parent parton, i.e. $d_{\mu \nu}(\wp,n)$, and dividing by $d-2$ (number of initial-state polarizations in $d=4-2\ep$ space-time dimensions). Explicitly, it is given by
\begin{align}
   \label{eq:gluonavg}
    \langle\hat{P}^{(0)}_{a_1\ldots a_m}\rangle = \frac{d_{\mu\nu}(\wt{P};n)}{d-2} \hat{P}^{(0), \mu\nu}_{a_1\ldots a_m}.
\end{align}

As we explained before, we shall choose a physical gauge to calculate the splitting amplitudes. Therefore, we use the light-cone gauge (LCG) where
\beq
d_{\mu \nu} (k,n) = - g_{\mu \nu} + \frac{k_{\mu} n_{\nu} + n_{\mu} k_{\nu}}{n\cdot k} \, ,
\label{bosonpol}
\eeq 
represents the physical polarization tensor of a gauge vector boson (gluon or photon) 
with momentum~$k$. The auxiliary gauge vector~$n$ in~\Eq{bosonpol} is 
taken identical to the light-like vector~$n$ introduced in \Eq{eq:Ptildegeneral}.

Finally, we compute the matrix elements associated with the evaluation of massive splitting matrices and splitting kernels by using two different frameworks. On the one side, we used \texttt{QGRAF}~\cite{Nogueira:1991ex} to generate all the necessary Feynman diagrams and an {\it in-house} code written in \texttt{Form}~\cite{Kuipers:2012rf} to perform Dirac and color algebra simplification. On the other, we relied on the automatized packages \texttt{FeynCalc}~\cite{Shtabovenko:2016sxi,Shtabovenko:2020gxv} and \texttt{FeynArts}~\cite{Hahn:2000kx}, exploiting their functions to simplify Dirac's chains. The results obtained are in total agreement within both methods. For the extraction of the splitting amplitudes and kernels, we follow the techniques described in refs.~\cite{Catani:1999ss,Catani:2002hc,Catani:2011st,Czakon:2022fqi}. In the following sections, we present perturbative results for various tree-level splitting processes up to $\mathcal{O}(\as^2)$ in the quasi-collinear limit described in eq.~(\ref{eq:quasicollimit}).

%***********************************************************************************************************************************************%
\section{Massive double-collinear splitting amplitudes and splitting kernels}
\label{sec:doublesplittings}

We recall in this section the known results for the double-collinear splitting matrices and kernels with massive quarks. There are two non-vanishing splitting processes, i.e.
\begin{align}
    \label{1to2processes}
    &Q\rar Q_1 g_2~,
    \\
    &g\rar Q_1 \bar{Q}_2~.
\end{align}
The remaining processes are related by charge conjugation invariance of QCD. Their corresponding splitting matrices are
\begin{align}
    \sp_{Q_1g_2}^{(0)} &= g_{\rm s}\mu_0^{\ep} \bt^{c_2}_{c_1 c}\fr{1}{\tilde{s}_{12}}\overline{u}(p_1)\slashed{\eps}(p_2)u(\wp)~,
    \\
    \sp_{Q_1\bar{Q}_2}^{(0)} &= g_{\rm s}\mu_0^{\ep}\bt^{c}_{c_1c_2}\fr{1}{s_{12}}\overline{u}(p_1)\slashed{\eps}^*(\wp)v(p_2)~,
\end{align}
where $\bt^c_{c_1 c_2}$ are the customary Gell-Mann colour matrices in the fundamental representation and the Dirac spinors are represented by $u(p)$ and $v(p)$. At the squared amplitude level, the splitting kernels are given by~\cite{Catani:2000ef,Keller:1998tf}
\begin{align}
    \hat{P}^{(0), ss^\prime}_{Q_1g_2} &= \delta^{ss^\prime}C_F\left[\frac{1+z^2}{1-z}-\ep(1-z)-2\fr{m_Q^2}{\tilde{s}_{12}} \right],
    \label{eq:PQgdoublemass}
    \\ \hat{P}^{(0), \mu \nu}_{Q_1\bar{Q}_2} &=T_R\left[-g^{\mu\nu}-4\frac{k_{\perp}^{\mu}k_{\perp}^{\nu}}{s_{12}} \right],
    \label{eq:PQQbdoublemass}
\end{align}
where $C_F = (N_c^2-1)/(2N_c)$ is the Casimir in the fundamental representation of SU($N_c$) QCD with $N_c$ colors. The corresponding Casimir in the adjoint representation is given by $C_A = N_c$ and $T_R=1/2$. The momentum fraction $z$ in eq.~(\ref{eq:PQgdoublemass}) is $z=z_1=1-z_2$ and the transverse momentum vector $k_{\perp}^{\mu}$ in eq.~(\ref{eq:PQQbdoublemass}) is $k_{\perp}^{\mu} = k_{\perp 1}^{\mu}=-k_{\perp 2}^{\mu}$.  As we anticipated in section \ref{sec:formalism}, the polarized splitting kernels are diagonal in the spin of the parent quark (anti-quark) but are not diagonal in the polarization of the parent gluon. To this end, we also obtain the corresponding unpolarized splitting kernels by using eqs.~(\ref{eq:quarkavg}-\ref{eq:gluonavg}) and they are as follows
\begin{align}
    \langle\hat{P}^{(0)}_{Q_1g_2}\rangle &= C_F\left[\frac{1+z^2}{1-z}-\ep(1-z)-2\fr{m_Q^2}{\tilde{s}_{12}} \right],
    \\ 
    \langle\hat{P}^{(0)}_{Q_1\bar{Q}_2}\rangle &=T_R\left[1-\frac{2}{1-\ep}\left(z(1-z)-\frac{m_Q^2}{s_{12}}\right) \right].
\end{align}
It is worth noticing that the mass dependence in $\hat{P}^{(0), \mu \nu}_{Q_1\bar{Q}_2}$ appears implicitly inside the sub-energy $s_{12}$; the functional structure is analogous to the massless case. In the case of $\hat{P}^{(0), ss^\prime}_{Q_1g_2}$, besides a contribution sharing the same functional dependence as in the massless case, there is an additional term proportional to $m^2_Q$. We will see similar features in the massive triple-collinear splitting kernels.

%***********************************************************************************************************************************************%
\section{Massive triple-collinear splitting amplitudes}
\label{sec:triplesplittings}
There are four non-vanishing splitting processes, which are given by
\begin{align}
    Q&\rar \bar{Q}_1 Q_2 Q_3~,
    \label{eq:4quarks}\\
    Q&\rar \bar{Q}_1^\prime Q_2^\prime Q_3~,
    \label{eq:twotwo}
    \\
    Q&\rar g_1 g_2 Q_3~,
    \\
    \label{eq:1to3gsplit}
    g&\rar g_1 Q_2 \bar{Q}_3~.
\end{align}
The quark-antiquark pair $\bar{Q}_1'Q_2'$ is of different flavour than $Q$ and has a mass $m_{Q'}$. All the other quarks and antiquarks have mass $m_Q$. In the following sections, we present results for the splitting matrices and splitting kernels corresponding to the processes in eqs.~(\ref{eq:4quarks}-\ref{eq:1to3gsplit}).

%%%%%%%%%%%%%%%%%%%%%%%%%%%%%%%%%%%%%%%%%%%%%%%%%%%%%%%%%%%%
\subsection{Splitting amplitudes with four massive quarks}
\label{ssec:splitting4quarks}
We consider first the splitting amplitude of four massive quarks of the same flavour, \Eq{eq:4quarks}. In this case there are two colour structures and they are related by the permutation $2\leftrightarrow 3$:
\begin{align}
    \mb{C}_1: T_R\left(\delta_{c_1c_3}\delta_{cc_2}-\fr{1}{N_c}\delta_{cc_3}\delta_{c_1c_2}\right)~,\\ \mb{C}_2: T_R\left(\delta_{c_1c_2}\delta_{cc_3}-\fr{1}{N_c}\delta_{cc_2}\delta_{c_1c_3}\right)~.
\end{align}
The l.h.s. of these expressions refer to operators, whilst the r.h.s. correspond to their projection according \Eq{eq:ColorDECO}. There are six distinct spin structures out of which three are related to the other three by the permutation $2\leftrightarrow 3$:
\begin{align}
    &\mb{S}_1: \bar{u}(p_3)\gamma^{\mu}u(\wp)\bar{u}(p_2)\gamma_{\mu}v(p_1)~, \quad \, \, \mb{S}_2: \bar{u}(p_3)\slashed{p}_2u(\wp)\bar{u}(p_2)\slashed{n}v(p_1)~, \\ &\mb{S}_3: \bar{u}(p_3)\slashed{p}_1 u(\wp)\bar{u}(p_2)\slashed{n}v(p_1)~,
    \qquad \mb{S}_4: \bar{u}(p_2)\gamma^{\mu}u(\wp)\bar{u}(p_3)\gamma_{\mu}v(p_1)~, \\ 
    &\mb{S}_5: \bar{u}(p_2)\slashed{p}_3 u(\wp)\bar{u}(p_3)\slashed{n}v(p_1)~, 
    \qquad \mb{S}_6: \bar{u}(p_2)\slashed{p}_1 u(\wp)\bar{u}(p_3)\slashed{n}v(p_1)~.
\end{align}
The splitting amplitude for the process is given by
\begin{align}
    \sp_{\bar{Q}_1 Q_2 Q_3}^{(0)} = \fr{g_{\rm s}^2\mu_0^{2\ep}}{s_{123}-m_{Q}^2}\left[\fr{\mb{C}_1}{s_{12}}\left(-\mb{S}_1+\fr{\mb{S}_2+\mb{S}_3}{n\cdot p_{12}} \right)-\fr{\mb{C}_2}{s_{13}}\left(-\mb{S}_4+\fr{\mb{S}_5+\mb{S}_6}{n\cdot p_{13}} \right)\right]~,
\end{align}
where $s_{123}\equiv(p_1+p_2+p_3)^2$ and $p_{ij}\equiv p_i+p_j$. If, instead, we consider the splitting process where the quark-antiquark pair $\bar{Q}_1' Q_2'$ is of different flavour and mass than the parent quark~$Q$, \Eq{eq:twotwo}, the splitting amplitude is simplified to
\begin{align}
    \sp_{\bar{Q}_1^\prime Q_2^\prime Q_3}^{(0)} = \fr{g_{\rm s}^2\mu_0^{2\ep}}{s_{123}-m_{Q}^2}\fr{\mb{C}_1}{s_{12}}\left(-\mb{S}_1+\fr{\mb{S}_2+\mb{S}_3}{n\cdot p_{12}} \right)~.
\end{align}
The explicit dependence on the mass $m_{Q^\prime}$ is hiding in $s_{123}$ and $s_{12}$, and the implicit dependence is through the momenta $p_1$ and $p_2$.

%%%%%%%%%%%%%%%%%%%%%%%%%%%%%%%%%%%%%%%%%%%%%%%%%%%%%%%%%%%%
\subsection{$Q\rar g_1 g_2 Q_3$}
\label{ssec:qg1g2Q3}
In this case there are two colour structures and they are related by the permutation $1\leftrightarrow 2$:
\begin{align}
   \mb{C}_1: \left(\bt^{c_2}\bt^{c_1}\right)_{c_3c}~,\quad \mb{C}_2: \left(\bt^{c_1} \bt^{c_2}\right)_{c_3c}~,
\end{align}
and eight distinct spin structures
\begin{align}
    &\mb{S}_1: \bar{u}(p_3)\slashed{p}_1\slashed{\eps}(p_1)\slashed{\eps}(p_2)u(\wp)~,\qquad\;\;\; \mb{S}_2: \bar{u}(p_3)\slashed{p}_2\slashed{\eps}(p_2)\slashed{\eps}(p_1)u(\wp)~,
    \nn\\
    &\mb{S}_3: \bar{u}(p_3)\slashed{\eps}(p_2)u(\wp)\,p_3\cdot \eps(p_1)~,\qquad\mb{S}_4: \bar{u}(p_3)\slashed{\eps}(p_2)u(\wp)\,p_2\cdot \eps(p_1)~,
    \nn\\
    &\mb{S}_5: \bar{u}(p_3)\slashed{\eps}(p_1)u(\wp)\,p_3\cdot \eps(p_2)~,\qquad\mb{S}_6: \bar{u}(p_3)\slashed{\eps}(p_1)u(\wp)\, p_1\cdot \eps(p_2)~,
    \nn\\
    &\mb{S}_7: \bar{u}(p_3)\slashed{p}_2u(\wp)\,\eps(p_1)\cdot \eps(p_2)~,\qquad \mb{S}_8: \bar{u}(p_3)\slashed{p}_1u(\wp)\,\eps(p_1)\cdot \eps(p_2)~.
\end{align}
Notice that these spin structures fulfil $\mb{S}_1\leftrightarrow \mb{S}_2$, $\mb{S}_3\leftrightarrow \mb{S}_5$, $\mb{S}_4\leftrightarrow \mb{S}_6$ and $\mb{S}_7\leftrightarrow \mb{S}_8$ under the transformation $1 \leftrightarrow 2$. The splitting amplitude for the process is given by
\begin{align}
    \sp^{(0)}_{g_1g_2Q_3} &= \fr{g_s^2\mu_0^{2\ep}}{\tilde{s}_{123}}\Bigg[\mb{C}_1\left(\fr{\mb{S}_2-2\mb{S}_5}{\tilde{s}_{23}} \right) 
    +\mb{C}_2\left(\frac{\mb{S}_1-2\mb{S}_3}{\tilde{s}_{13}}\right) \nn \\
&  + (\mb{C}_2 - \mb{C}_1 )\fr{2}{\tilde{s}_{12}}\left(\mb{S}_4+\frac{n\cdot p_2}{n\cdot p_{12}}\mb{S}_8 -\mb{S}_6-\frac{n\cdot p_1}{n\cdot p_{12}}\mb{S}_7\right)
    \Bigg]~,
\end{align}
being the full expression symmetric under the exchange $1 \leftrightarrow 2$, as expected. Here, we define $\tilde{s}_{123}= \tilde{s}_{12}+ \tilde{s}_{13}+\tilde{s}_{23}$, following the notation introduced in eq. (\ref{eq:pijmassive}).

%%%%%%%%%%%%%%%%%%%%%%%%%%%%%%%%%%%%%%%%%%%%%%%%%%%%%%%%%%%%
\subsection{$g\rar g_1 Q_2 \bar{Q}_3$}
\label{ssec:gg1Q2Qb3}
In this case there are two colour structures given by:
\begin{align}
   \mb{C}_1: \left(\bt^{c}\bt^{c_1}\right)_{c_2c_3}~,\qquad \mb{C}_2: \left(\bt^{c_1}\bt^{c}\right)_{c_2c_3}~,
\end{align}
and six distinct spin structures
\begin{align}
    &\mb{S}_1: \bar{u}(p_2)\slashed{p}_1\slashed{\eps}(p_1)\slashed{\eps}^*(\wp)v(p_3)~, \qquad\quad \mb{S}_2: \bar{u}(p_2)\slashed{\eps}^*(\wp)v(p_3)\,p_2\cdot \eps(p_1)~,
    \nn\\
    &\mb{S}_3: \bar{u}(p_2)\slashed{\eps}^*(\wp)v(p_3)\,p_3\cdot \eps(p_1)~, \quad\quad\,\mb{S}_4: \bar{u}(p_2)\slashed{\eps}(p_1)v(p_3)\,p_1\cdot \eps^*(\wp)~,
    \nn\\
    &\mb{S}_5: \bar{u}(p_2)\slashed{p}_1v(p_3)\,\eps^*(\wp)\cdot \eps(p_1)~,\quad\quad\,\mb{S}_6: \bar{u}(p_2)\slashed{n}v(p_3)\, \eps^*(\wp)\cdot \eps(p_1)~.
    \label{eq:BasisgqQQb}
\end{align}
However, these spin structures can be grouped according to
\beq
\mb{S}'_1 = \mb{S}_1 -  2\mb{S}_2 \, , \quad \mb{S}'_2 = 2 \, (\mb{S}_2 + \mb{S}_3 + \mb{S}_4 - \mb{S}_5) \, ,
\eeq
in such a way that the splitting amplitude for the process is written as
\begin{align}
    \sp^{(0)}_{g_1Q_2\bar{Q}_3} &=\fr{g_s^2\mu_0^{2\ep}}{s_{123}}\Bigg[ \mb{C}_1\left(\fr{\mb{S}'_1+\mb{S}'_2}{\tilde{s}_{13}} \right) +\mb{C}_2 \, \fr{\mb{S}'_1}{\tilde{s}_{12}} + (\mb{C}_1- \mb{C}_2)
    \left( \fr{\mb{S}'_2}{s_{23}}+\fr{s_{123}}{s_{23}}\fr{\mb{S}_6}{n\cdot p_{23}}\right) 
    \Bigg].
\end{align}
It is worth appreciating that, in the massless limit, we recover the same expressions presented in ref.~\cite{Czakon:2022fqi}: in fact, we explicitly defined the spinor bases to ease the comparison and check our results. Besides, we appreciate that both the massive and massless splitting amplitudes share the same spinor and colour structures.

%***********************************************************************************************************************************************%
\section{Massive triple-collinear splitting kernels}
\label{sec:SquaredTriple}

In order to obtain compact expressions, we exploit the results provided in ref.~\cite{Catani:1999ss} for the massless triple-collinear kernels, including the definition of the kinematical variable
\beq
t_{ij,k} = 2 \frac{z_i \tilde{s}_{jk}-z_j \tilde{s}_{ik}}{z_i+z_j} + \frac{z_i -z_j}{z_i+z_j} \tilde{s}_{ij} \, .
\label{eq:Variabletijk}
\eeq

The unpolarized splitting kernel involving two quark-antiquark pairs of different flavours is given by
\begin{align}
\label{eq:Q2QpbQpQ}
    \la\hat{P}^{(0)}_{\bar{Q}_1^\prime Q_2^\prime Q_3}\ra  &= C_F T_R\Bigg\{\fr{\tilde{s}_{12}\tilde{s}_{123}}{2s_{12}^2} \left[-\fr{t_{12,3}^2}{\tilde{s}_{12}\tilde{s}_{123}}+\fr{4z_3+(z_1-z_2)^2}{1-z_3}+(1-2\ep)\left(z_1+z_2-\fr{\tilde{s}_{12}}{\tilde{s}_{123}}\right)\right]
    \nn\\
    &+\fr{2m_{Q^\prime}^2}{s_{12}^2}\bigg[\fr{z_3 \tilde{s}_{123} }{(1-z_3)^2}(1+2z_3-3z_3^2+4z_1z_2)-\fr{\tilde{s}_{23}}{1-z_3}(2-3z_1-5z_2+z_1^2+z_2^2)  \nn\\
    &-\fr{\tilde{s}_{13}}{1-z_3}(2-5z_1-3z_2+z_1^2+z_2^2)-\ep\Big(\tilde{s}_{123}(1-z_3)-\tilde{s}_{12}(1+z_3)\Big)\bigg]-2\fr{m_{Q}^2\tilde{s}_{12}}{s_{12}^2}
    \nn\\
    &+\fr{4m_{Q^\prime}^4}{s_{12}^2}z_3\left[\ep+\fr{2z_1z_2}{(1-z_3)^2}+\fr{2z_3}{1-z_3}\right]-4\fr{m_{Q}^2m_{Q^\prime}^2}{s_{12}^2}\Bigg\}~.
\end{align}

The splitting kernel involving two quark-antiquark pairs of the same flavour is decomposed as
\begin{align}
    \la\hat{P}^{(0)}_{\bar{Q}_1 Q_2 Q_3}\ra  =\left[\la\hat{P}^{(0)}_{\bar{Q}_1^\prime Q_2^\prime Q_3}\ra\Big|_{m_{Q^\prime}=m_{Q}}+(2\leftrightarrow 3)\right] +\la\hat{P}_{\bar{Q}_1 Q_2 Q_3}^{\rm{(0,id)}}\ra~,
\end{align}
where the interference contribution is given by
\begin{align}
    \la\hat{P}_{\bar{Q}_1 Q_2 Q_3}^{\rm{(0,id)}}\ra &= C_F \left(C_F -\fr{C_A}{2}\right)\Bigg\{\fr{\tilde{s}_{12}\tilde{s}_{13}}{s_{12}s_{13}}\Bigg[ (1-\ep)\left(\fr{2\tilde{s}_{23}}{\tilde{s}_{12}}-\ep \right)+\fr{\tilde{s}_{123}}{\tilde{s}_{12}}\bigg\{\fr{1+z_1^2}{1-z_2}-\fr{2z_2}{1-z_3}
    \nn\\
    &-\ep\left(\fr{(1-z_3)^2}{1-z_2}+1+z_1-\fr{2z_2}{1-z_3}\right)-\ep^2(1-z_3)\bigg\}-\fr{\tilde{s}_{123}^2}{\tilde{s}_{12}\tilde{s}_{13}}\fr{z_1}{2}\bigg\{\fr{1+z_1^2}{(1-z_2)(1-z_3)}
    \nn\\
    &-\ep\left(1+2\fr{1-z_2}{1-z_3}\right)-\ep^2\bigg\}\Bigg]+\fr{m_{Q}^2}{s_{12}s_{13}}\Bigg[2\tilde{s}_{23}\fr{3-7z_2+z_2^2+z_2^3+4z_2^2z_3}{(1-z_2)(1-z_3)}
    \nn\\
    &+2\tilde{s}_{12}\fr{(1-2z_2)^2+(1-2z_3)^2-z_3(1-z_3^2)-z_2(1+2z_3)+z_2z_3(2z_2+3z_3)}{(1-z_2)(1-z_3)} 
    \nn\\
    &-2\ep \tilde{s}_{12}\fr{2(1-z_2)^2+(1-z_3)^2-z_3(1-z_3^2)-2z_2z_3(1+z_1)}{(1-z_2)(1-z_3)}
    \nn\\
    &-2\ep \tilde{s}_{23}\fr{1-z_2^2+z_2(1-3z_3)}{1-z_3}+\ep^2\Big(2\tilde{s}_{12}(1-z_3)-\tilde{s}_{23}(1-z_1)\Big)\Bigg] 
    \nn\\
    &+\fr{2m_Q^4}{s_{12}s_{13}}\Bigg[2\fr{(1-2z_2)^2-z_2(1-z_2z_3)}{(1-z_2)(1-z_3)}-\ep\fr{(1-2z_2)(3-2z_2-z_2z_3)}{(1-z_2)(1-z_3)}+\ep^2 \Bigg]\Bigg\}
    \nn\\
    &+(2\leftrightarrow 3)~.
\end{align}

The splitting kernel involving two gluons, and a massive quark as parent parton, admits the following decomposition
\begin{align}
    \la\hat{P}^{(0)}_{g_1 g_2 Q_3}\ra  =C_F^2\la\hat{P}_{g_1 g_2 Q_3}^{\rm{(0,ab)}}\ra +C_A C_F\la\hat{P}_{g_1 g_2Q_3}^{\rm{(0,nab)}}\ra~,
\end{align}
where
\begin{align}
    \la\hat{P}_{g_1 g_2 Q_3}^{\rm{(0,ab)}}\ra&=\Bigg\{\fr{\tilde{s}_{123}^2z_3}{2\tilde{s}_{13}\tilde{s}_{23}}\left[\fr{1+z_3^2}{z_1 z_2}-\ep\fr{z_1^2+z_2^2}{z_1 z_2}-\ep(1+\ep)\right]+\fr{\tilde{s}_{123}}{\tilde{s}_{13}}\bigg[\fr{z_3(1-z_1)+(1-z_2)^3}{z_1 z_2}
    \nn\\
    &-\ep(z_1^2+z_1z_2+z_2^2)\fr{1-z_2}{z_1z_2}+\ep^2(1+z_3) \bigg]+(1-\ep)\left[\ep-(1-\ep)\fr{\tilde{s}_{23}}{\tilde{s}_{13}}\right]
    \nn\\
    &+\fr{m_{Q}^2}{\tilde{s}_{23}}\bigg[\frac{2(z_2-2z_3+z_2^2+2z_3^2+3z_2z_3)}{z_1z_2}-\fr{2(2z_2+4z_3-z_1z_3)}{z_1}
    \nn\\
    &-\fr{2\tilde{s}_{123}}{\tilde{s}_{23}}\fr{1+z_2^2+z_3^2+2z_2z_3}{z_1}-\fr{\tilde{s}_{12}}{\tilde{s}_{13}}\fr{z_2(1-z_3)+4z_3}{z_2}+2\ep z_1\fr{\tilde{s}_{123}}{\tilde{s}_{23}}
    \nn\\
    &+\ep(1-z_3)\fr{\tilde{s}_{123}}{\tilde{s}_{13}}\bigg]+\fr{4m_Q^4}{\tilde{s}_{23}^2}+\fr{4m_Q^4}{\tilde{s}_{13}\tilde{s}_{23}}\Bigg\} + (1\leftrightarrow 2)~,
\end{align}
and
\begin{align}
\label{eq:PggQnab}
    \la\hat{P}_{g_1 g_2 Q_3}^{\rm{(0,nab)}}\ra&=\Bigg\{ (1-\ep)\left(\fr{t_{12,3}^2}{4\tilde{s}_{12}^2}+\fr{1}{4}-\fr{\ep}{2}\right) +\fr{\tilde{s}_{123}^2}{2\tilde{s}_{12}\tilde{s}_{13}}\bigg[\fr{(1-z_3)^2(1-\ep)+2z_3}{z_2}
    \nn\\
    &+\fr{z_2^2(1-\ep)+2(1-z_2)}{1-z_3}\bigg]-\fr{\tilde{s}_{123}^2}{4\tilde{s}_{13}\tilde{s}_{23}}z_3\left[\fr{(1-z_3)^2(1-\ep)+2z_3}{z_1z_2}+\ep(1-\ep)\right]
    \nn\\
    &+\fr{\tilde{s}_{123}}{2\tilde{s}_{12}}\left[(1-\ep)\fr{z_1(2-2z_1+z_1^2)-z_2(6-6z_2+z_2^2)}{z_2(1-z_3)}+2\ep\fr{z_3(z_1-2z_2)-z_2}{z_2(1-z_3)}\right]
    \nn\\
    &+\fr{\tilde{s}_{123}}{2\tilde{s}_{13}}\bigg[(1-\ep)\fr{(1-z_2)^3+z_3^2-z_2}{z_2(1-z_3)}-\ep\left(\fr{2(1-z_2)(z_2-z_3)}{z_2(1-z_3)}-z_1+z_2\right)
    \nn\\
    &-\fr{z_3(1-z_1)+(1-z_2)^3}{z_1z_2}+\ep(1-z_2)\left(\fr{z_1^2+z_2^2}{z_1z_2}-\ep\right)\bigg]-\fr{2m_Q^2}{\tilde{s}_{12}}\fr{z_1^2+z_2^2}{z_1z_2}
    \nn\\
    &+\fr{m_Q^2}{\tilde{s}_{13}\tilde{s}_{23}}\left[\fr{\tilde{s}_{12}}{2}\fr{z_2+4z_3-z_2z_3}{z_2}+\tilde{s}_{13}\fr{2z_2+4z_3-z_1z_3}{z_1}-\fr{\ep}{2}\tilde{s}_{123}(1-z_3)\right]
    \nn\\
    &-\fr{m_Q^2}{\tilde{s}_{12}\tilde{s}_{23}}\bigg[\tilde{s}_{12}\fr{3z_1z_2^2+z_1^2z_2-2z_1^2z_3+2z_2^2z_3+4z_2^3}{z_1z_2(1-z_3)}
    \nn\\
    &+2(\tilde{s}_{13}z_2-\tilde{s}_{23}z_1)\fr{z_1^2+z_1z_2+z_2^2}{z_1z_2(1-z_3)}\bigg]-\fr{2m_Q^4}{\tilde{s}_{13}\tilde{s}_{23}}\Bigg\} + (1\leftrightarrow 2)~.
\end{align}
Finally, the splitting kernel with a gluon as a parent parton is decomposed as
\begin{align}
    \hat{P}^{(0),\mu\nu}_{g_1Q_2\bar{Q}_3} = C_FT_R\hat{P}^{\rm{(0,ab)},\mu\nu}_{g_1Q_2\bar{Q}_3} + C_AT_R\hat{P}^{\rm{(0,nab)},\mu\nu}_{g_1Q_2\bar{Q}_3}~,
\end{align}
where
\begin{align}
  \hat{P}^{\rm{(0,ab)},\mu\nu }_{g_1Q_2\bar{Q}_3} &= g^{\mu\nu}\bigg[2-\fr{2\tilde{s}_{123}\tilde{s}_{23}+(1-\ep)(\tilde{s}_{123}-\tilde{s}_{23})^2}{\tilde{s}_{12}\tilde{s}_{13}}+2m_{Q}^2\left(\frac{\tilde{s}_{123}}{\tilde{s}_{12}^2}+\frac{\tilde{s}_{123}}{\tilde{s}_{13}^2}-\frac{2\tilde{s}_{23}}{\tilde{s}_{12}\tilde{s}_{13}}\right)
  \nn\\
  &+4m_Q^4\left(\fr{1}{\tilde{s}_{12}^2}+\fr{1}{\tilde{s}_{13}^2}\right)\bigg]+\fr{4\tilde{s}_{123}}{\tilde{s}_{12}\tilde{s}_{13}}\left[\tilde{k}_2^{\mu}\tilde{k}_3^{\nu}+\tilde{k}_3^{\mu}\tilde{k}_2^{\nu}-(1-\ep)\tilde{k}_1^{\mu}\tilde{k}_1^{\nu}\right] 
  \nn\\ &+8m_Q^2\left[\fr{\tilde{k}_2^{\mu}\tilde{k}_2^{\nu}}{\tilde{s}_{13}^2}+\fr{\tilde{k}_3^{\mu}\tilde{k}_3^{\nu}}{\tilde{s}_{12}^2}\right]-\fr{8m_Q^2}{\tilde{s}_{12}\tilde{s}_{13}}(1-\ep)\tilde{k}_1^{\mu}\tilde{k}_1^{\nu}~,
\end{align}
and
\begin{align}
    \hat{P}^{\rm{(0,nab)},\mu\nu }_{g_1Q_2\bar{Q}_3} &= \Bigg\{-\fr{2g^{\mu\nu}}{s_{23}^2}\bigg[\fr{z_3\tilde{s}_{12}(z_2\tilde{s}_{13}-z_3\tilde{s}_{12})}{(1-z_1)^2}+\fr{z_2(1-2z_2)(2-2z_2-5z_3)}{2z_1(1-z_1)^2}\tilde{s}_{23}^2 
    \nn\\
    &+\fr{\tilde{s}_{12}\tilde{s}_{23}}{z_1(1-z_1)^2}\Big(z_2+z_3-2z_2^2-3z_3^2-3z_2z_3+2z_2^3+3z_3^3+5z_2^2z_3+6z_2z_3^2 \Big)\bigg]
    \nn\\
    &-\fr{2}{s_{23}^2}\bigg[\left(\fr{4z_3^2\tilde{s}_{123}}{z_1(1-z_1)}-\fr{2(1-z_2)\tilde{s}_{23}}{z_1}\right)\kttm\kttn-\left(\fr{4z_2z_3\tilde{s}_{123}}{z_1(1-z_1)}-\fr{\tilde{s}_{23}(1-z_1)}{z_1}\right)\kttm\kthtn
    \nn\\
    &+\ep \tilde{s}_{23}\kotm\kotn\bigg]-4g^{\mu\nu}\fr{m_Q^2}{s_{23}^2}\bigg[ \fr{2z_2-4z_2^2-3z_2z_3}{(1-z_1)^2}\tilde{s}_{23}+\tilde{s}_{12}\bigg(\fr{z_3(z_2-z_3)}{(1-z_1)^2}+\fr{z_1}{1-z_1}
    \nn\\
    &+\fr{1-z_1}{z_1}\bigg)\bigg]+\fr{4m_Q^2}{s_{23}^2}\bigg[\! \left(\!2+\fr{z_3(z_2-z_3)}{z_1(1-z_1)}\!\right)2\kttm\kttn+\left(\!2-\fr{(z_2-z_3)^2}{z_1(1-z_1)}\right)\kttm\kthtn-\ep\kotm\kotn\bigg]
    \nn\\
    &+4g^{\mu\nu}\fr{m_Q^4}{s_{23}^2}\left[1-\fr{z_1}{1-z_1}-\fr{z_2z_3}{(1-z_1)^2}\right]+\fr{g^{\mu\nu}}{2s_{23}}\bigg[\fr{2-3z_1}{1-z_1}\fr{\tilde{s}_{23}^2}{\tilde{s}_{13}}-\fr{z_3(1-2z_1)}{z_1(1-z_1)}\fr{\tilde{s}_{12}^2}{\tilde{s}_{13}}
    \nn\\
    &+\fr{z_2\tilde{s}_{13}}{z_1(1-z_1)}-\fr{1-2z_2}{z_1}\tilde{s}_{12}-\fr{1-3z_2-4z_3+2z_2^2+5z_3^2+7z_2z_3}{z_1(1-z_1)}\fr{\tilde{s}_{12}\tilde{s}_{23}}{\tilde{s}_{13}}
    \nn\\
    &-\fr{1-2z_2-z_3+3z_2^2+4z_3^2+7z_2z_3}{z_1(1-z_1)}\tilde{s}_{23}\bigg]+\fr{1}{s_{23}}\bigg[\left(\fr{\tilde{s}_{123}}{\tilde{s}_{13}}-\fr{1-z_3}{z_1}\right)2\kthtm\kthtn
    \nn\\
    &+\left(\fr{z_2-z_3-z_2^2+z_3^2+2z_2z_3}{z_1(1-z_1)}\fr{\tilde{s}_{123}}{\tilde{s}_{13}}-\fr{z_3}{z_1}\fr{\tilde{s}_{23}}{\tilde{s}_{13}}+\fr{1-z_1}{z_1}\right)\left(\kttm\kthtn+\kthtm\kttn\right)
    \nn\\
    &+\left(\fr{1-z_2}{z_1}\fr{\tilde{s}_{23}-\tilde{s}_{13}}{\tilde{s}_{13}}-\fr{2z_3^2}{z_1(1-z_1)}\fr{\tilde{s}_{123}}{\tilde{s}_{13}}\right)2\kttm\kttn+\ep\bigg( \fr{\tilde{s}_{123}}{\tilde{s}_{13}}\left(\kotm\kthtn+\kthtm\kotn\right)
    \nn\\
    &+\fr{\tilde{s}_{23}}{\tilde{s}_{13}}\left(\kotm\kttn+\kttm\kotn\right)+2\kotm\kotn\bigg)\bigg]+\fr{2m_Q^2}{s_{23}\tilde{s}_{13}}\bigg[\left(3+\fr{z_3(z_2-z_3)}{z_1(1-z_1)}\right)2\kttm\kttn+2\kthtm\kthtn
    \nn\\
    &+\left(\fr{z_3(z_2-z_3)}{z_1(1-z_1)}+\fr{3z_2+z_3}{1-z_1}\right)\left(\kttm\kthtn+\kthtm\kttn\right)-2\ep \kotm\kotn\bigg]+2g^{\mu\nu}\fr{m_Q^4}{s_{23}\tilde{s}_{13}}\fr{2-3z_1}{1-z_1}
    \nn\\
    &-g^{\mu\nu}\fr{m_Q^2}{s_{23}}\bigg[\fr{z_1^2+2z_3-z_1z_2-4z_1z_3}{z_1(1-z_1)}\fr{\tilde{s}_{12}}{\tilde{s}_{13}}+\fr{z_1^2+2z_2^2+4z_3^2+z_1z_3+6z_2z_3}{z_1(1-z_1)}
    \nn\\
    &-\fr{2(2-3z_1)}{1-z_1}\fr{\tilde{s}_{23}}{\tilde{s}_{13}}\bigg]+g^{\mu\nu}\bigg[\fr{\tilde{s}_{23}^2}{2\tilde{s}_{12}\tilde{s}_{13}}+\fr{z_1-2z_2}{4z_1}\fr{\tilde{s}_{23}}{\tilde{s}_{12}}-\fr{z_2}{4z_1}\fr{\tilde{s}_{13}}{\tilde{s}_{12}}-\fr{z_3}{4z_1}\fr{\tilde{s}_{12}}{\tilde{s}_{13}}+\fr{1-z_1}{4z_1}
    \nn\\
    &+\fr{z_1-2z_3}{4z_1}\fr{\tilde{s}_{23}}{\tilde{s}_{13}}-\fr{\ep}{2}\bigg]+\left[\fr{1-z_2}{z_1}\fr{\tilde{s}_{23}}{\tilde{s}_{12}\tilde{s}_{13}}+\fr{1-z_2}{z_1\tilde{s}_{12}}-\fr{z_3}{z_1}\fr{\tilde{s}_{123}}{\tilde{s}_{12}\tilde{s}_{13}}\right]\kttm\kttn
    \nn\\
    &+\left[\fr{\tilde{s}_{123}}{\tilde{s}_{12}\tilde{s}_{13}}-\fr{1-z_3}{z_1\tilde{s}_{12}}\right]\kthtm\kthtn+\left[\fr{1-2z_3}{2z_1}\fr{\tilde{s}_{23}}{\tilde{s}_{12}\tilde{s}_{13}}+\fr{1-2z_2}{2z_1}\fr{\tilde{s}_{123}}{\tilde{s}_{12}\tilde{s}_{13}}+\fr{z_2-z_3}{z_1\tilde{s}_{12}}\right]\kttm\kthtn
    \nn\\
    &+\left[\fr{1}{2z_1}\fr{\tilde{s}_{12}+\tilde{s}_{13}}{\tilde{s}_{12}\tilde{s}_{13}}-\fr{\tilde{s}_{123}}{\tilde{s}_{12}\tilde{s}_{13}}\right]\kthtm\kttn+\ep\left[\fr{\kthtm\kotn}{\tilde{s}_{13}}+\fr{\kotm\kttn}{\tilde{s}_{12}}-\fr{\tilde{s}_{23}}{\tilde{s}_{12}\tilde{s}_{13}}\kotm\kotn\right]
    \nn\\
    &+g^{\mu\nu}m_Q^2\left[\fr{\tilde{s}_{23}}{\tilde{s}_{12}\tilde{s}_{13}}-\fr{z_1+2z_2}{2z_1\tilde{s}_{12}}-\fr{z_1+2z_3}{2z_1\tilde{s}_{13}}\right]+\fr{2m_Q^2}{\tilde{s}_{12}\tilde{s}_{13}}\bigg[\kttm\kthtn-\kthtm\kttn-\kotm\kthtn
    \nn\\
    &-\kttm\kotn-\ep \kotm\kotn\bigg]\Bigg\} + (2\leftrightarrow 3).
\end{align}
By following the description given in eq.~(\ref{eq:gluonavg}), which is equivalent to performing an azimuthal average, we obtain the unpolarized gluon-initiated splitting kernel. It is given by
\begin{align}
    \la\hat{P}^{(0)}_{g_1Q_2\bar{Q}_3}\ra = C_FT_R \, \la\hat{P}^{\rm{(0,ab)}}_{g_1Q_2\bar{Q}_3}\ra + C_AT_R\, \la\hat{P}^{\rm{(0,nab)}}_{g_1Q_2\bar{Q}_3}\ra~,
\end{align}
with
\beqn
\nonumber \la\hat{P}^{\rm{(0,ab)}}_{g_1Q_2\bar{Q}_3}\ra &=& \Bigg\{\left[ \frac{s_{123}^2 \left(z_1^2 (1-\epsilon )+2 (1-z_2) z_3-\epsilon \right)}{\tilde{s}_{12} \tilde{s}_{13}(1-\epsilon)} \right.
\\ \nonumber &+& \left.\frac{2 s_{123} ((z_1+1) \epsilon +z_2-1)}{\tilde{s}_{12}(1-\epsilon)} + \frac{\tilde{s}_{13} (1-\epsilon )}{\tilde{s}_{12}}-\epsilon \right]
\\ \nonumber &+& \left[ \frac{m_Q^2}{\tilde{s}_{12}} \,\left(\frac{2 s_{123} (2 (1-z_3) z_3+\epsilon -1)}{\tilde{s}_{12}}+\frac{2 s_{123} (z_1+2 z_2 z_3+\epsilon )}{\tilde{s}_{13}}-4 \right) \right.
\\ &-& \left.  \,\frac{4 \, m_Q^4}{\tilde{s}_{13}} \left(\frac{1}{\tilde{s}_{12}}+\frac{1}{\tilde{s}_{13}}\right) \right]\frac{1}{1-\epsilon}\Bigg\} \,   + (2\leftrightarrow 3) \, ,
\eeqn
and
\beqn
\label{eq:g2gQQpNab}
\nonumber \la\hat{P}^{\rm{(0,nab)}}_{g_1Q_2\bar{Q}_3}\ra &=&  \Bigg\{\left[  \frac{s_{123}^2 z_3 }{2 s_{23} \tilde{s}_{13}}\left(\frac{(1-z_1)^3-z_1^3}{z_1 (1-z_1)}-\frac{2 z_3 (-2 z_1 z_2-z_3+1)}{z_1 (1-z_1) (1-\epsilon )}\right) \right.
\\ \nonumber &-& \left.  \frac{s_{123}^2 }{2 \tilde{s}_{12} \tilde{s}_{13}}\left(z_1^2-\frac{z_1+2 z_2 z_3}{1-\epsilon }+1\right) -\frac{t_{23,1}^2}{4 s_{23}^2}+\frac{\epsilon }{2}-\frac{1}{4}\right. 
\\ \nonumber &+& \left. \frac{s_{123} }{2 s_{23}}\left(\frac{z_1^3+1}{z_1 (1-z_1)}+\frac{z_1 (z_3-z_2)^2-2 (z_1+1) z_2 z_3}{z_1 (1-z_1) (1-\epsilon )}\right) \right.   
\\ \nonumber &+& \left.  \frac{s_{123} (1-z_2) }{2 \tilde{s}_{13}}\left(-\frac{2 (1-z_2) z_2}{z_1 (1-z_1) (1-\epsilon )}+\frac{1}{z_1 (1-z_1)}+1\right) \right] 
\\ \nonumber &+& \left[ \frac{m_Q^2}{s^2_{23}} \,\left(\frac{2 s_{123}^2 (1-z_2)}{\tilde{s}_{13} (1-z_1) z_1}-\frac{s_{123}^3 (z_1+2 z_2 z_3+ \epsilon )}{\tilde{s}_{12} \tilde{s}_{13}}-\frac{2 s_{123}^2 z_1^2 (1-2 z_2)}{\tilde{s}_{13} (1-z_1)} \right. \right.
\\ \nonumber &+& \left. \left. \frac{2 s_{123}^2(4 (1-z_2) z_2+z_2+2 \epsilon -2)}{\tilde{s}_{13}}-\frac{2 s_{123} \tilde{s}_{12} z_2}{\tilde{s}_{13} (1-z_1)}-\frac{2 s_{123} \tilde{s}_{12} z_3}{\tilde{s}_{13} z_1} \right. \right.
\\ \nonumber &-& \left. \left. s_{123} (z_1 (1-4 z_2)+4 (1-z_2) z_2+2 \epsilon +3)+\frac{s_{123}}{z_1} + 4 \tilde{s}_{12} \right. \right.
\\ \nonumber &-& \left. \left.  \frac{2 s_{123} \tilde{s}_{12} \left(z_2-2 z_2^2+\epsilon \right)}{\tilde{s}_{13}} - (1-\epsilon) \frac{2 z_2 (s_{23}+2 \tilde{s}_{13}) (z_2-z_3)}{(1-z_1)^2} \right)  \right. \, 
\\ &+&   \left. m_Q^4 \, \left(\frac{2}{\tilde{s}_{12}\tilde{s}_{13}} + \frac{(1-\epsilon)(z_2-z_3)^2}{s_{23}^2 (1-z_1)^2}\right) \right] \frac{1}{1-\epsilon}\Bigg\} \,   + (2\leftrightarrow 3) \, . 
\eeqn
As an independent and highly non-trivial check on our results presented in eqs.~(\ref{eq:Q2QpbQpQ}-\ref{eq:g2gQQpNab}), in the massless limit, we confirm the expressions for the triple-collinear splitting kernels given in ref.~\cite{Catani:1999ss}. These expressions are also available in a machine-readable \texttt{Mathematica} file on \texttt{Zenodo} \cite{ZENODO}.

Finally, let us comment on the extension of the present results to other theories, in particular to QED. All the expressions presented here were computed within QCD, but we carefully stripped the color factors. Besides, in $g\to g Q \bar{Q}$ and $Q\to g g Q$, we explicitly separated the splitting kernels into Abelian and non-Abelian contributions. Thus, we can compute the corresponding QED splitting kernels by replacing gluons by photons, taking the Abelian part and replacing the color factors by electric charges. For the four-quark splittings, the QCD results are already Abelian (i.e. gluon self-interactions are absent), so we have
\beqn
\la\hat{P}^{\rm{(0,QED)}}_{\bar{Q}_1^\prime Q_2^\prime Q_3}\ra &=& e_Q^2 e_{Q^\prime}^2 \left[\la\hat{P}^{\rm{(0)}}_{\bar{Q}_1^\prime Q_2^\prime Q_3}\ra\right]_{C_A\to 0,\,C_F \to 1,\,T_R \to 1} \, ,
\\ \la\hat{P}^{\rm{(0,QED)}}_{\bar{Q}_1 Q_2 Q_3}\ra &=& e_Q^4 \left[\la\hat{P}^{\rm{(0)}}_{\bar{Q}_1 Q_2 Q_3}\ra\right]_{C_A\to 0,\,C_F \to 1,\,T_R \to 1} \, ,
\eeqn
whilst for the remaining cases, we write
\beqn
\la\hat{P}^{\rm{(0,QED)}}_{\gamma_1 \gamma_2 Q_3}\ra &=& e_Q^4 \la\hat{P}^{\rm{(0,ab)}}_{g_1 g_2 Q_3}\ra \, ,
\\
\hat{P}^{\rm{(0,QED)}}_{\gamma_1 Q_2\bar{Q}_3} &=& e_Q^4 \hat{P}^{\rm{(0,ab)}}_{g_1 Q_2\bar{Q}_3} \, .
\eeqn
These expressions are in agreement with the relations presented in eq. (71) from ref. \cite{Catani:1999ss} for the massless case. The procedure described before is known in general as {\it Abelianization} \cite{deFlorian:2015ujt,deFlorian:2016gvk,deFlorian:2018wcj,Ajjath:2019ixh,Ajjath:2019vmf}, and it is independent on the mass of the particles.
%***********************************************************************************************************************************************%
\section{Conclusions and outlook}
\label{sec:conclusions}
In this article, we have presented the first calculation of triple-collinear splitting amplitudes and (un)polarized splitting kernels with massive partons at tree-level in QCD. We have carefully compared the kinematics of the collinear and quasi-collinear limits, and have provided very compact explicit expressions for all the possible processes involving QCD partons.

Although collinear singularities are strictly absent for splitting processes with massive particles, mass-dependent logarithmically enhanced terms emerge in scattering amplitudes from the quasi-collinear region. Since these contributions depend on ratios of masses to the typical energy scale of the process, they could lead to large corrections in the high-energy limit. So, in order to improve theoretical predictions it is necessary to use subtraction counter-terms involving massive splitting kernels. At NLO, the massive dipole formalism~\cite{Catani:2002hc} required double-collinear splittings to build the corresponding counter-terms. The massive triple-collinear splittings provided in this article allow to extend the collinear subtraction formalism to reach NNLO accuracy, taking into account parton mass effects. This might play a crucial role in collider phenomenology involving heavy quarks. For instance, in forward charm-quark production \cite{Maciula:2022lzk}, there are no complete calculations that keeps track of the mass of the charm quark, which is expected to have an important phenomenological impact.

Also, having access to the full set of triple-collinear splitting kernels with massive partons paves the road for more precise parton shower generators. Up to now, their improvement relied mostly on the inclusion of higher-logarithmic corrections, since mass effects were always consider as sub-dominant. However, as we are reaching next-to-next-to-leading logarithmic accuracy or even higher-orders, mass effects start to compete and can no longer be neglected.

Finally, the knowledge of triple-collinear splitting processes with massive partons could play an important role in the consistent treatment of mass effects in parton densities evolution. For this purpose, higher-order corrections to the Altarelli-Parisi kernels of splitting functions, keeping the exact mass dependence, and generalising the DGLAP equations to the quasi-collinear limit are required. The results provided in this work constitute a very first step towards this direction, and might be relevant in sight of the current precision high-energy collider phenomenology program. An independent calculation of the quark-initiated massive triple-collinear splitting kernels has recently been presented in ref.~\cite{Craft:2023aew}, which is in agreement with our result.
%%%%%%%%%%%%%%%%%%%%%%%%%%%%%%%%%%%%%%%%%%%%%%%%%%%%%%%%%%%%%%%%%%%%%%%%%%%%%%%%%%%%%%
%%%%%%%%%%%%%%%%%%%%%%%%%%%%%%%%%%%%%%%%%%%%%%%%%%%%%%%%%%%%%%%%%%%%%%%%%%%%%%%%%%%%%%
\section*{Acknowledgements}
We warmly thank the Galileo Galilei Institute (Firenze), where this work was completed, and S. Catani, for hospitality and discussing during the Workshop {\it Theory Challenges in the Precision Era of the Large Hadron Collider}, A. Szczurek and M. Wiesemann for their fruitful comments when presenting this work at EPS-HEP 2023, and the authors of ref.~\cite{Craft:2023aew} for a careful comparison of the quark-initiated massive triple-collinear splitting kernels.
This work is supported by the Spanish Government (Agencia Estatal de Investigaci\'on MCIN /AEI/10.13039/501100011033) Grants No. PID2020-114473GB-I00, PID2022-141910NB-I00 and Generalitat Valenciana Grant No. PROMETEO/2021/071. 
The work of PKD is supported by European Commission MSCA Action COLLINEAR-FRACTURE, Grant Agreement No. 101108573.
The work of GS is partially supported by EU Horizon 2020 research and innovation programme STRONG-2020 project under Grant Agreement No. 824093 and H2020-MSCA-COFUND USAL4EXCELLENCE-PROOPI-391 project under Grant Agreement No 101034371.

%%%%%%%%%%%%%%%%%%%%%%%%%%%%%%%%%%%%%%%%%%%%%%%%%%%%%%%%%%%%%%%%%%%%%%%%%%%%%%%%%%%%%%
%%%%%%%%%%%%%%%%%%%%%%%%%%%%%%%%%%%%%%%%%%%%%%%%%%%%%%%%%%%%%%%%%%%%%%%%%%%%%%%%%%%%%%

\bibliographystyle{JHEP}

\providecommand{\href}[2]{#2}\begingroup\raggedright\endgroup

\end{document}